\documentclass{elsart}

\usepackage{graphicx}
\usepackage{amssymb}

\begin{document}
\begin{frontmatter}
\title{Jaggedness of path integral trajectories\thanksref{MNTR}}
\thanks[MNTR]{Supported by the Ministry of
Science and Environmental Protection of the Republic of
Serbia through projects No. 1486 and 1899.}

\author{A. Bogojevi\' c\corauthref{alex}},
\ead{alex@phy.bg.ac.yu}
\corauth[alex]{Corresponding author.}
\author{A. Bala\v z}, \and
\author{A. Beli\' c}
\address{Scientific Computing Laboratory, Institute of Physics\\
P. O. Box 57, 11001 Belgrade, Serbia and Montenegro}

\begin{abstract}
We define and investigate the properties of the jaggedness of path integral trajectories. The new quantity is shown to be scale invariant and to satisfy a self-averaging property. Jaggedness allows for a classification of path integral trajectories according to their relevance. We show that in the continuum limit the only paths that are not of measure zero are those with jaggedness 1/2, i.e. belonging to the same equivalence class as random walks. The set of relevant trajectories is thus narrowed down to a specific subset of non-differentiable paths. For numerical calculations, we show that jaggedness represents an important practical criterion for assessing the quality of trajectory generating algorithms. We illustrate the obtained results with Monte Carlo simulations of several different models.
\end{abstract}

\begin{keyword}
Path integral \sep Quantum theory \sep Self-similarity \sep Monte Carlo
\PACS 03.65.-w \sep 03.65.Db \sep 05.45.Df \sep 05.10.Ln 
\end{keyword}
\end{frontmatter}

\section{Introduction}
\label{sec:intro}

The set of all the trajectories that one integrates over in a path integral is huge (cardinality $\aleph_2$). In order to have successful numerical simulations it is important to generate a finite number of trajectories that are as representative of the whole set as possible. For this reason it is necessary to broaden the amount of analytical information regarding the subset of trajectories that give dominant contributions to path integrals. This subset is much smaller than the set of all trajectories. Investigation of its properties is of great importance. Ultimately we would like to find ways to generate only these relevant trajectories. 

Classification of trajectories according to their relevance is just as important for analytical calculations of continuum limit amplitudes. As a result of the continuum limit many trajectories are in fact of measure zero, i.e. give no contribution to path integrals. For example, it is well known that differentiable trajectories are of measure zero. This has been a severe impediment to the development of path integration theory as most of mathematics (and practically all of our intuition) is firmly based on smooth, differentiable functions. In fact, the principle reason that we have made any analytical progress with path integrals lies in the fact that, although of measure zero, some differentiable functions (e.g. classical solutions) turn out to be important markers of ``nearby" non-differentiable paths that do give important contributions. 

We are in a precarious position in which we know very little about the paths we need to work with. In addition, most of our knowledge is negative, i.e. tells us which trajectories are not important rather than which are. This is a substantial problem in the analytical approach. In numerical simulations, on the other hand, we work with discrete trajectories about whose classification we know even less. Therefore, the implementing of a systematic classification of the relevance of path integral trajectories (both in the discretized and continuum theories) is a crucial starting point in the quest for more efficient calculation schemes.

One of the few positive statements concerning path integral trajectories is that they are stochastically self-similar \cite{feynmanhibbs,feynman}. An exhaustive review of various aspects of the path integral method can be found in \cite{kleinert}. A direct consequence of self-similarity is that relevant path integral trajectories have \cite{kroeger} fractal (Hausdorff) dimension $d_H=2$. In this paper we introduce a new classification property -- the jaggedness of a trajectory. An analytical and numerical analysis of the properties of jaggedness leads to a new classification of trajectories according to their relevance to path integrals.

\section{Self-similarity of trajectories}
\label{sec:selfsim}

The property of stochastic self-similarity of path integral trajectories has important repercussions both on the dynamics as well as on the construction of efficient numerical algorithms for generating paths. In a previous set of papers \cite{prl05a,prb05a,pla05a} we have used self-similarity to obtain analytical relations between discretizations of different coarseness for the case of a general theory. The newly developed analytical method systematically improves the convergence of
path integrals of a generic $N$-fold discretized theory through the explicit construction of a set of effective actions $S^{(p)}$ for $p=1,2,3,\ldots$. These effective actions are equivalent to the starting action (in the sense that they lead to the same continuum amplitudes), however, the path integrals calculated using them converge to the continuum limit ever faster. Discretized amplitudes calculated using the $p$ level effective action tend to the continuum limit as $1/N^p$. Using the general procedure we obtained explicit effective actions up to $p=9$. Self-similarity played a crucial role in this procedure allowing us to derive, and asymptotically solve, an integral equation relating discretized theories viewed at different coarseness. It was shown \cite{pla05a} that this approach is in fact equivalent to the derivation of a generalization of Euler's summation formula to path integrals.

Self-similarity has also been successfully utilized in constructing efficient numerical algorithms for generating paths for Monte Carlo simulations \cite{pollockceperley,ceperley}. In particular the L\'evy construction discussed in these references generates self-similar paths through a simple iterative procedure. One begins with the fixed end points $a$ at time $t=0$ and $b$ at time $t=T$ and samples a bisecting point at time $t=T/2$ from a Gaussian centered in the middle between $a$ and $b$ and with width $\sigma=\sqrt{T/2}$. Having sampled the point at $t=T/2$ one now bisects the two new intervals $[0,T/2]$ and $[T/2,T]$ in the same way generating new bisection points at $t=T/4$ and $t=3T/4$ with appropriately centered Gaussians of width $\sigma=\sqrt{T/4}$. The procedure is continued recursively doubling the number of sampled points at each level and using the width $\sigma=\sqrt{\epsilon/2}$, where $\epsilon$ is the current time step. This method exactly samples free particle trajectories, but also works very well for interacting theories. The principle benefit of the method is that computational effort scales as $O(N)$, where $N$ is the coarseness of the discretization.

It is important to strengthen the relation between these two types of approaches. For a deeper understanding of the role of self-similarity it is necessary to classify trajectories according to their relative contribution to the path integral. As we have already mentioned, the first and most natural such classification of paths was with respect to their fractal dimension. The fractal dimension $d_H$ is calculated from the formula 
\begin{equation}
\label{fractaldimension}
\langle L\rangle\propto N^{\frac{d_H-1}{d_H}}\ ,
\end{equation}
where $\langle L\rangle$ is the expectation value of $L=\sum_{i=0}^{N-1}|q_{i+1}-q_i|$, the total trajectory length. It was shown by Kr\" oger \cite{kroeger} that for theories with (Euclidean) action of the form
\begin{equation}
\label{action}
S=\int dt\,\left(\frac{1}{2}\dot q^2+V(q)\right)\ ,
\end{equation}
the only trajectories that contribute to the path integral are those with fractal dimension $d_H=2$. The reason for this is that for short times of propagation the kinetic term dominates over the potential and each model looks like a random walk (which has $d_H=2$). This is illustrated in Fig. 1 on the example of an anharmonic oscillator with quartic coupling $V(q)=\frac{1}{2}q^2+\frac{1}{4!}gq^4$. Kr\" oger has also shown that the addition of velocity dependent interactions changes the fractal dimension of relevant paths.
\begin{figure}[!ht]
\centering
\includegraphics[height=4.8cm]{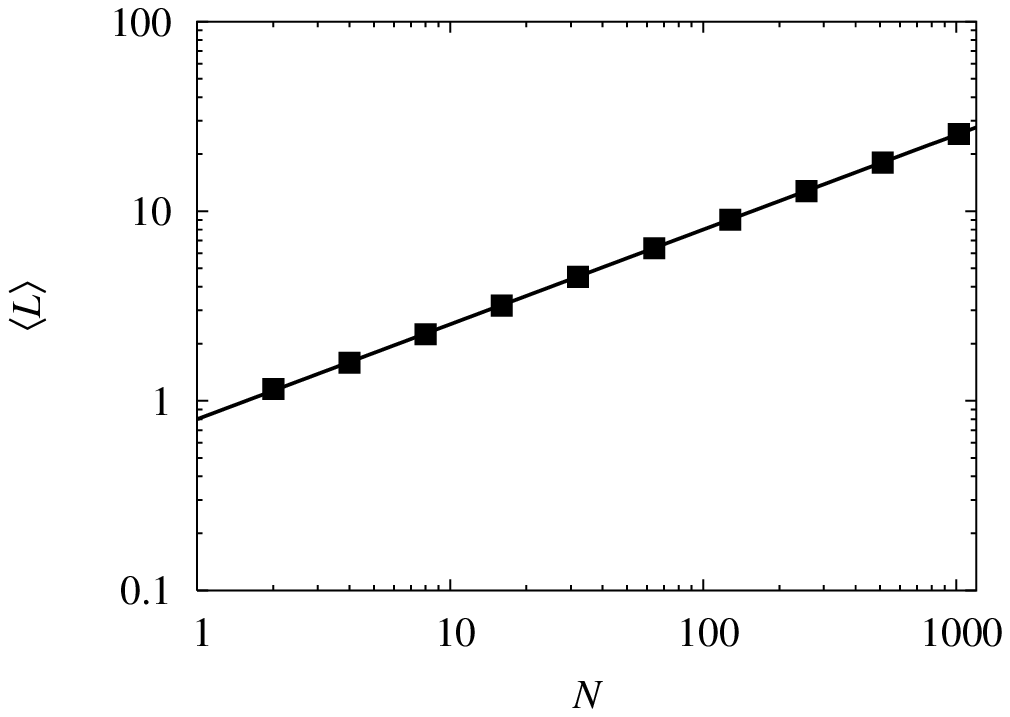}
\includegraphics[height=4.8cm]{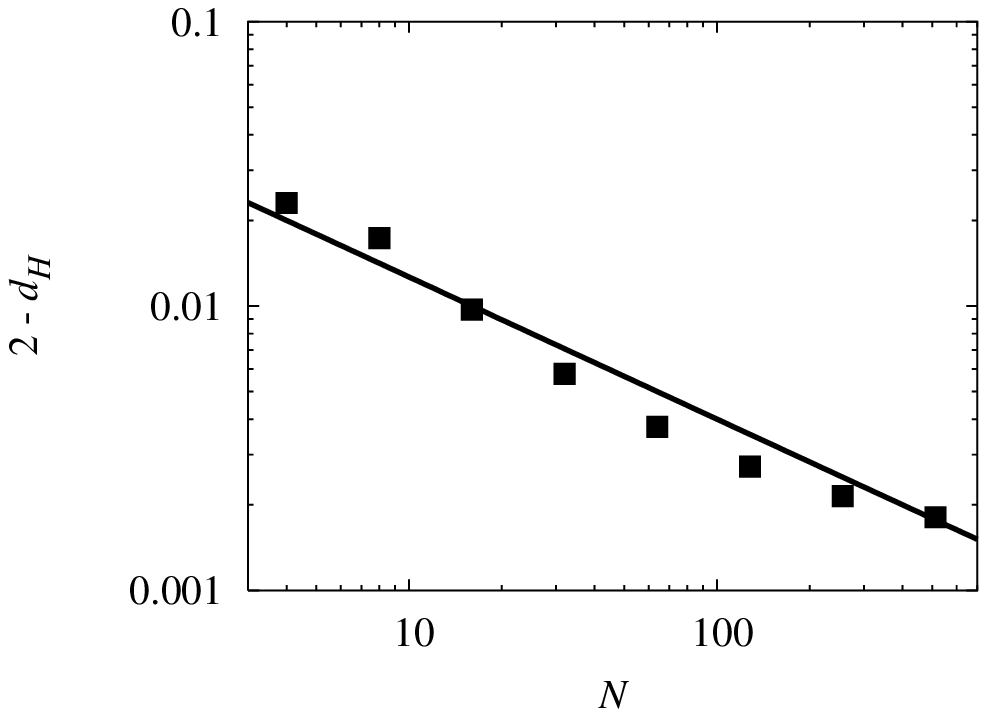}
\caption{(Left) A typical graph of the expectation value of the trajectory length $\langle L\rangle$ as a function of discretization coarseness $N$. The data fits  to $\langle L\rangle=0.8\sqrt{N}$, i.e. to $d_H=2$. (Right) A detailed picture of how the fractal dimension approaches the value for a random walk in the continuum limit fits to $2-d_H=0.04/\sqrt{N}$. Both plots correspond to propagation from $a=0$ to $b=1$ in time $T=1$ for an anharmonic oscillator with quartic coupling $g=1$. The number of Monte Carlo samples used was $N_{MC}=9.2\cdot 10^6$.}
\end{figure}
\begin{figure}[!ht]
\centering
\includegraphics[width=10cm]{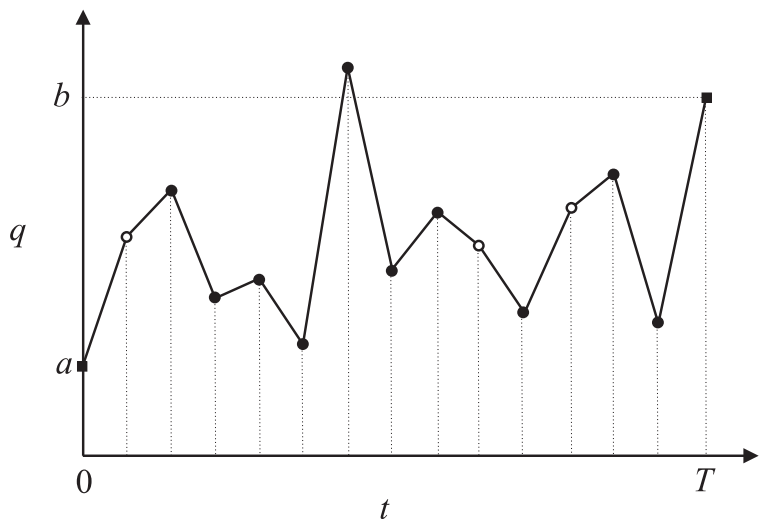}
\caption{An example of a trajectory that goes from $a$ (at $t=0$) to $b$ (at $t=T$) in $N=14$ discrete time steps.}
\end{figure}
Let us now introduce another quantity (complementary to the fractal dimension) which will serve to further classify path integral trajectories according to their relevance. 

The $N$-fold discretized path from $q_0=a$ to $q_N=b$ is determined by the $N-1$ intermediate positions $q_1, q_2, \ldots, q_{N-1}$. We define the jaggedness of a path as
\begin{equation}
\label{jaggedness}  
J=\frac{1}{N-1}\sum_{i=0}^{N-2}\frac{1}{2}\Big(1-\mbox{sgn}(\delta_i\delta_{i+1})\Big)\ ,
\end{equation}
where $\delta_i=q_{i+1}-q_i$. 
From the above definition we see that $J$ in fact counts the number of peaks (both minima and maxima) divided by $N-1$ (the maximal number peaks for an $N$-fold discretized trajectory). Therefore, for all values of $N$, we have $J\in[0,1]$. Fig. 2 illustrates a typical path with $N=14$ and jaggedness $J=\frac{10}{13}$ 
(intermediate points that are peaks are depicted by black circles).

An important property of the jaggedness is that it is scale invariant up to $1/N$ terms. Namely, if we scale the intermediate points $q_i\to\lambda q_i$ the only thing that can change are the first and last term in the above sum (since the end points need to remain fixed). This brings about a change in $J$ that is $O(1/N)$. Therefore, we see that in the continuum limit $J$ is in fact scale invariant. $J$ is also scale invariant for finite $N$ when $a=b=0$. This behavior is quite different from that of fractal dimension. From eq. (\ref{fractaldimension}) we see that the fractal dimension is not invariant under scaling with an $N$-dependent $\lambda$. For this reason the  classification of trajectories with respect to $J$ is independent of a classification with respect to $d_H$. 

\section{Analytical and numerical analysis of jaggedness}
\label{sec:jag}

In the continuum limit all smooth and differentiable trajectories have $J=0$. This is not only true of monotonic trajectories but also of those having a finite number of extrema. Paths with $J=0$ are of measure zero. A classification of paths with non-vanishing jaggedness is therefore a path integral motivated classification of non-differentiable paths.

From eq. (\ref{jaggedness}) we see that (up to $1/N$ terms) the jaggedness satisfies an averaging property, i.e. if we split a trajectory with $2N$ discrete time steps into two equal halves (with jaggedness $J_1$ and $J_2$), the total jaggedness equals
\begin{equation}
J=\frac{1}{2}(J_1+J_2)-\frac{1}{2N-1}\left[\frac{1-\mbox{sgn}(\delta_{N-1}\delta_N)}{2}-\frac{J_1+J_2}{2}\right]\ .
\end{equation}
As we can see, for finite $N$ the averaging is broken by a $1/N$ term. In the continuum limit, however, the averaging of jaggedness is exact. Cutting up a trajectory into $k$ equal pieces we get $J=\frac{1}{k}\sum_{i=1}^kJ_i$. Stochastic self-similarity now implies that all the $J_i$'s are equal to each other, and in fact that they are equal to the jaggedness of the whole path. We denote this property of jaggedness as self-averaging. The jaggedness of the whole trajectory (corresponding to a motion for time $T$) is thus equal to the jaggedness of even the smallest piece of that trajectory (corresponding to the propagation for a much shorter time $\Delta t$). 

It is well known that for short times of propagation the dynamics of any model is well approximated by that of a random walk, i.e. does not depend on the potential. Therefore, self-similarity of trajectories implies that the jaggedness of the trajectories of a general model with action of the form given in eq. (\ref{action}) is equal to the jaggedness of trajectories for a random walk. 

The expectation value of the jaggedness for a random walk is quite easily calculated. For a random walk $\delta_i$ and $\delta_{i+1}$ are not correlated and so $\langle J\rangle =\frac{1}{2}$. We therefore conclude that for all the models with action of the form of eq. (\ref{action}) we have $\langle J\rangle=\frac{1}{2}$. Similarly, for finite $N$ we find that $\langle J\rangle$ deviates from the continuum as $O(1/N)$. 

\begin{figure}[!ht]
\centering
\includegraphics[width=10cm]{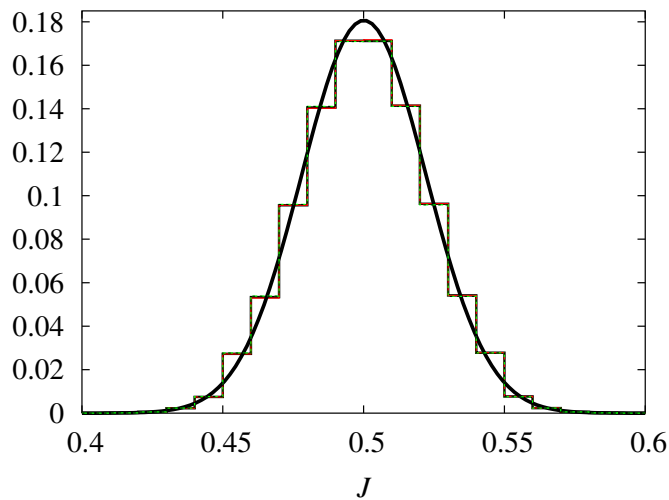}
\caption{Relative contributions to the path integral of trajectories of different jaggedness for different models. The histograms are practically indistinguishable for: anharmonic oscillators with quartic coupling $g=1$ (black) and $g=1000$ (red) as well as particles in a modified P\"oschl-Teller potential $V(q)=-\frac{\alpha^2}{2}\frac{\beta(\beta-1)}{\cosh^2(\alpha q)}$ with parameters $\alpha=0.5$, $\beta=1.5$ (green) and $\alpha=0.5$, $\beta=2$ (dashed black). The data is for $N=512$, $T=1$, $a=0$, $b=1$, $N_{MC}=9.2\cdot 10^6$ and fits to a Gaussian centered at $0.5$ with width $0.022$. The number of bins is 100.}
\end{figure}
\begin{figure}[!ht]
\centering
\includegraphics[width=10cm]{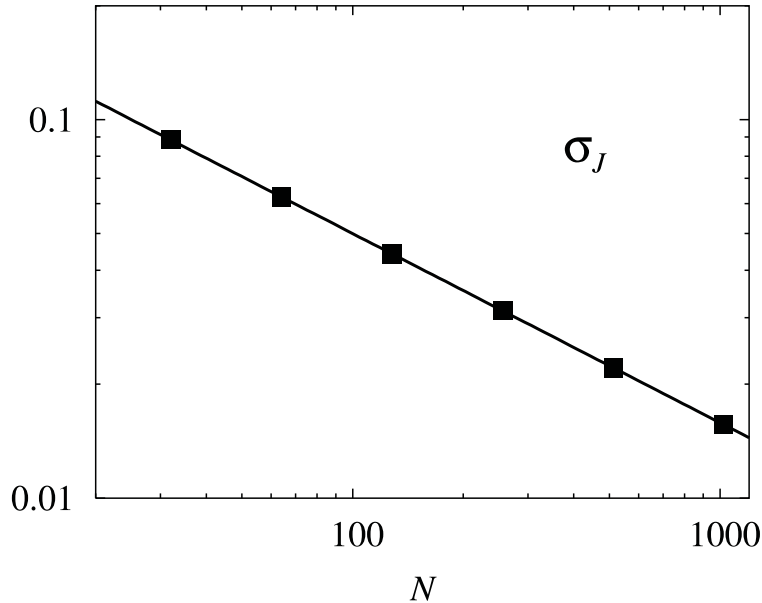}
\caption{Width of the distribution around $\langle J\rangle$ plotted as a function of $N$. The data fits to $\sigma_J=0.5/\sqrt{N}$. The plot is for an anharmonic oscillator with quartic coupling $g=1$, propagating for $T=1$ from $a=0$ to $b=1$ and $N_{MC}=9.2\cdot 10^6$.}
\end{figure}

Fig. 3 gives the relative contributions to the path integral of trajectories of different jaggedness for the case of $N=512$. As is shown in the figure, the distribution is practically indistinguishable for different models and widely varying parameters. Detailed numerical investigations show that for all models studied deviations from a Gaussian distribution (e.g. skew and kurtosis) go to zero as $N\to\infty$. In agreement with our previous analytical argument based on self-similarity the center of the distribution is $\langle J\rangle = \frac{1}{2}+ O(1/N)$.  
The width of the distribution $\sigma_J$ vanishes as $\sigma_J\sim 1/\sqrt{N}$ as illustrated in Fig. 4. The conclusion is that for models with action of the form of eq. (\ref{action}) the only paths that are not of measure zero are those with $J=\frac{1}{2}$. 

There are several important consequences of this that we need to mention. The analytical consequence of the above results is that they give us a much more detailed understanding of which trajectories are relevant and which are not. Not only are smooth, differentiable trajectories of measure zero (i.e. those with $J=0$), but in fact most of the non-differentiable trajectories also do not contribute to path integrals. To understand path integrals better we need to focus on a much narrower class of non-differentiable trajectories -- those with $J=\frac{1}{2}$, i.e. those belonging to the same jaggedness equivalence class as the random walk. 

The numerical consequence of the above results is that for finite $N$ it is important to have algorithms that generate trajectories that are distributed in a way that mimics, as much as possible, the physical distribution of relative contributions to the path integral such as the one shown in Fig. 3. The most important thing to focus on is the center of the distributions.  
\begin{figure}[!ht]
\centering
\includegraphics[width=10cm]{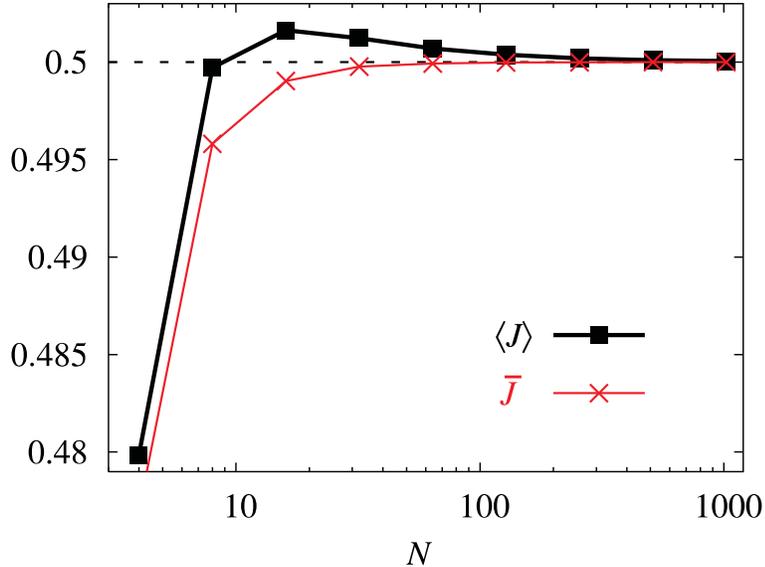}
\caption{$N$-dependances of $\langle J\rangle$ and of $\bar J$. The fact that $\bar J$ tends to the same continuum value as the physical expectation value indicates that the Le\' vy construction is a good algorithm to use for large $N$. Parameters are the same as in Fig. 4.}
\end{figure}

Fig. 5 compares the physical average $\langle J\rangle$, which is algorithm independent, with $\bar J$ the average jaggedness of the trajectories generated by the algorithm (in this case the Le\' vy construction). Note that for the Le\' vy construction $\bar J$ is completely independent of the dynamics. The reason why this method represents a good algorithm (at least for large $N$) is that $\bar J\to\frac{1}{2}$ in the continuum limit. This figure also indicates that a useful measure of the quality $Q$ of an algorithm for generating trajectories is roughly the inverse of $|\langle J\rangle - \bar J|$. We will use this measure of quality to compare the Le\' vy construction with the diagonalization algorithm in which trajectories are generated by a Gaussian distribution function using a semi-classical expansion. The computing time of this algorithm scales as $O(N^2)$ since it is necessary to diagonalize the quadratic form in the exponential of the distribution function.

In Fig. 6 we plot $|\langle J\rangle - \bar J|$ for these two algorithms. The top curve is for the Le\' vy construction, the bottom for the diagonalization method. The latter algorithm outperforms the former by an order of magnitude for all values of $N$. As a result of this the Monte Carlo errors calculated using the two methods differ by an order of magnitude. However, one should not forget that quality needs to be balanced by an assessment of cost (in computing time). Although it gives a relatively lower quality, the computing time for the Le\' vy method scales as $N$. The diagonalization method outperforms the Le\' vy method for a given $N$, however its computation time scales as $N^2$. Both algorithms are good (since $|\langle J\rangle - \bar J|$ vanishes in the continuum limit). A simple cost-benefit analysis tells us that Le\' vy construction is better for large $N$, while diagonalization is better for smaller $N$.
\begin{figure}[!ht]
\centering
\includegraphics[width=10cm]{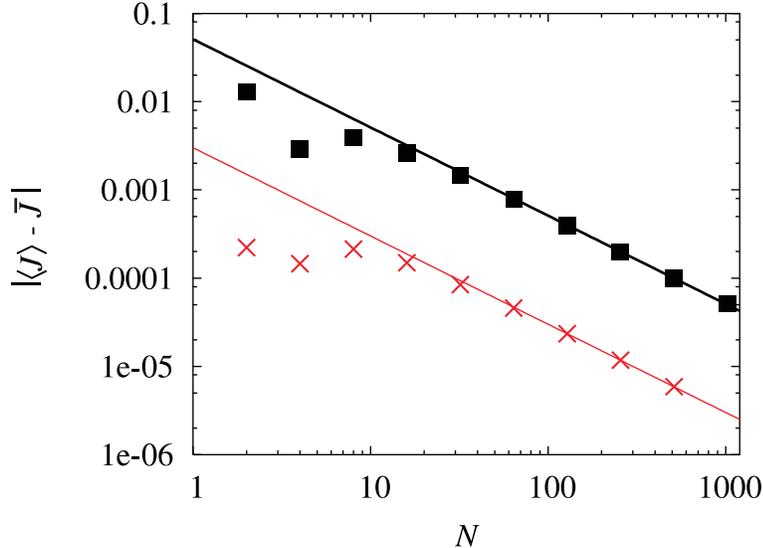}
\caption{Comparison of $\langle J\rangle$ and $\bar J$ calculated using two different algorithms for the case of an anharmonic oscillator with quartic coupling. The top curve was calculated using the Le\' vy bisection method. The data fits to the curve $|\langle J\rangle - \bar J|_{Levy}=0.051/N$. The lower curve was calculated using the diagonalization method. The data fits to $|\langle J\rangle - \bar J\,|_{diag}=0.003/N$.}
\end{figure}

Not all algorithms for generating paths are good, however. For example, if we modify the Le\' vy construction using uniform distributions (of appropriate widths) instead of Gaussians we get an algorithm for which $\bar J$ does not tend to $\frac{1}{2}$. A signature of a bad algorithm is that there exists a maximum quality $Q_{max}$ that can't be passed irrespective of computational cost. 

As we have mentioned, it is important to strengthen the connection between the dynamical (calculation of path integrals) and kinematical (generation of trajectories) aspects of self-similarity. The effective action approach \cite{prl05a,prb05a,pla05a} has brought about an immense speedup in path integral calculations. The speedup is a direct consequence of the fact that effective actions allow us to work with much smaller values of $N$ to obtain the same precision. Using standard algorithms (which have all been derived to be optimal for large $N$) we have obtained a speedup of many orders of magnitude. The important step that next needs to be taken is to develop a path generating algorithm that is tailored for small $N$'s, i.e. for which $\bar J$ is near to $\langle J\rangle$ for coarse discretizations. Note that the diagonalization algorithm is one such method, but that it is not computationally optimized. We are currently working on developing these kinds of algorithms and tying them in to the effective action approach. 

At the very end we wish to make further contact between jaggedness and random walks. In order to be near the random walk limit, the potential term in the discretized action $\epsilon V$, where $\epsilon=T/N$, must be smaller than the kinetic term $\delta^2/2\epsilon$. One should expect that when we are near the random walk regime all quantities depend on the ratio of these two. On the other hand, for a random walk we have $\delta^2/\epsilon\sim 1$, so that the ratio is in fact $\epsilon V(q_c)$ where $q_c$ is a characteristic length. A rough value for the characteristic length follows from $q_c\, p\sim 1$ (essentially Heisenberg's uncertainty relation in $\hbar=1$ units). We are using units in which $m=1$, so $p=\delta/\epsilon\sim1/\sqrt{\epsilon}$. The last step follows from the basic random walk relation $\delta^2/\epsilon\sim 1$. Finally we find that everything should be expressed in terms of the ratio $\epsilon V(\sqrt{\epsilon})$. For example, for the anharmonic oscillator with quartic coupling this ratio is $\epsilon^3g$. We have shown that $\langle J\rangle$ differs from $\frac{1}{2}$ by a term proportional to $1/N$. From this it follows that for the oscillator with quartic anharmonicity $|\langle J\rangle -\frac{1}{2}|$ should be proportional to $g^{1/3}/N$. A similar back of the envelope calculation for a particle moving in a modified P\"oschl-Teller potential gives that $|\langle J\rangle -\frac{1}{2}|$ should be proportional to $\frac{\alpha^2\beta(\beta-1)}{N}$. Fig. 7 illustrates that these simple calculations do in fact hold.
\begin{figure}[!ht]
\centering
\includegraphics[height=4.8cm]{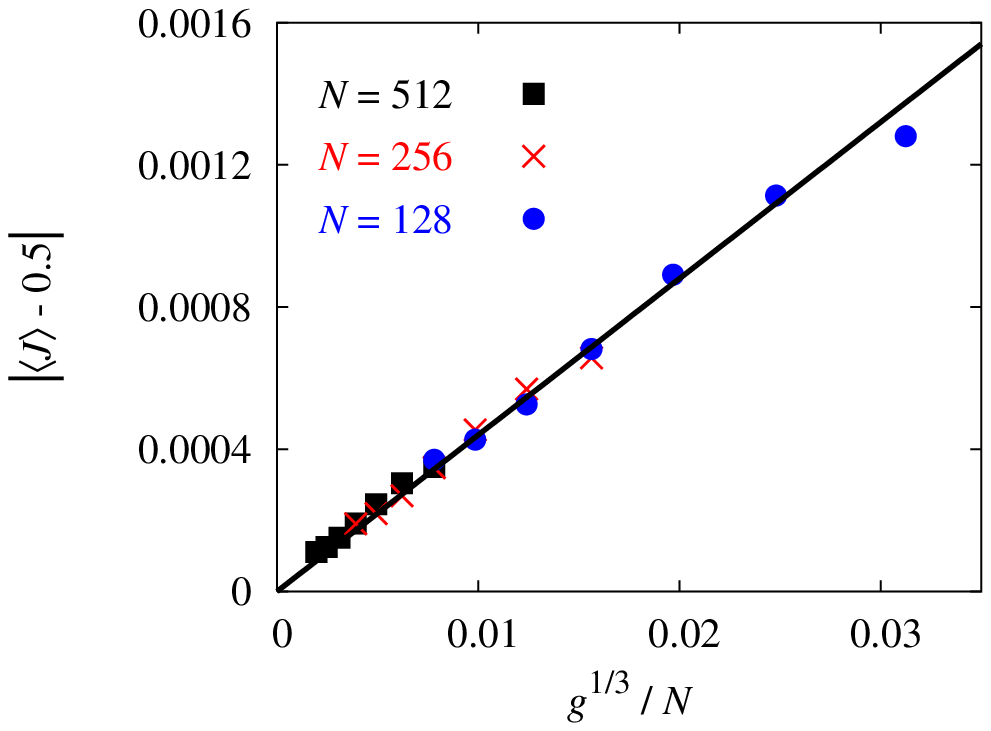}
\includegraphics[height=4.8cm]{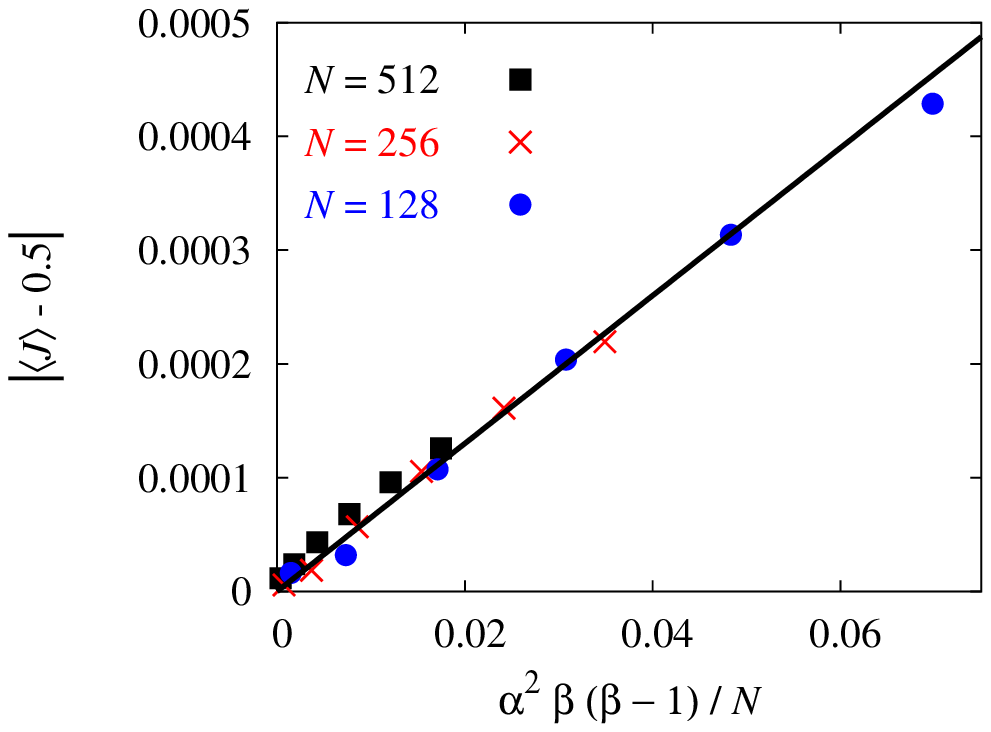}
\caption{(Left) Anharmonic oscillators with quartic coupling $g=1,2,4,8,\ldots,64$ for $N=128,256$ and $512$. (Right) Modified P\"oschl-Teller potential for $\alpha=0.5$ and $\beta=1.5,2.5,\ldots,6.5$ for $N=128,256$ and $512$. All the simulations were done for $a=0$, $b=1$, $T=1$ and $N_{MC}=9.2\cdot 10^6$.}
\end{figure}

To conclude, we have identified and investigated the properties of a quantity that we call the jaggedness and that is useful for obtaining a more detailed classification of relevant path integral trajectories. For discrete calculation, i.e. numerical simulations, the properties of the jaggedness are useful for obtaining more efficient algorithms for generating representative paths. Furthermore, we have shown that jaggedness can be used as an important practical criterion of the quality of trajectory generating algorithms. In the continuum limit (analytical calculations) we found that only trajectories with jaggedness equal to $\frac{1}{2}$ contribute to the path integral. In this way we greatly narrow the set of trajectories that are not of measure zero to those belonging to the equivalence class of the random walk. Classification of paths with respect to jaggedness is thus a classification of relevant non-differentiable paths. In the continuum limit, jaggedness is shown to be scale invariant as well as self-averaging.


\begin{thebibliography}{00}

\bibitem{feynmanhibbs}
R. P. Feynman and A. R. Hibbs,
\emph{Quantum Mechanics and Path Integrals}
(McGraw-Hill, New York, 1965).

\bibitem{feynman}
R. P. Feynman,
\emph{Statistical Mechanics}
(W. A. Benjamin, New York, 1972).

\bibitem{kleinert}
H. Kleinert,
\emph{Path Integrals in Quantum Mechanics, Statistics, Polymer Physics, and Financial Markets}
(World Scientific, 2004).

\bibitem{kroeger}
H. Kr\" oger,
Phys. Rep. {\bf 323}, 81 (2000). 

\bibitem{prl05a}
A. Bogojevi\'c, A. Bala\v z, and A. Beli\'c, 
Phys. Rev. Lett. {\bf 94}, 180403 (2005).

\bibitem{prb05a}
A. Bogojevi\'c, A. Bala\v z, and A. Beli\'c, 
Phys. Rev. B {\bf 72}, 064302 (2005).

\bibitem{pla05a}
A. Bogojevi\'c, A. Bala\v z, and A. Beli\'c, 
Phys. Lett. A {\bf 344}, 84 (2005).

\bibitem{pollockceperley}
E. L. Pollock and D. M. Ceperley,
Phys. Rev. B {\bf 30}, 2555 (1984).

\bibitem{ceperley}
D. M. Ceperley,
Rev. Mod. Phys. {\bf 67}, 279 (1995).

\end{thebibliography}
\end{document}